\documentclass[aps,apl,twocolumn,superscriptaddress,a4paper]{revtex4}

\usepackage{graphicx}
\usepackage{dcolumn}
\usepackage{bm}
\usepackage{amsmath}

\begin{document}
\DeclareGraphicsExtensions{.pdf}


\title{Multi-layer atom chips for versatile atom micro manipulation}

\author{M. Trinker}
 \affiliation{Atominstitut, Vienna University of Technology, 1020 Vienna, Austria}
\author{S. Groth}
 \affiliation{Physikalisches Institut, Universit\"at Heidelberg, 69120 Heidelberg, Germany}
\author{S. Haslinger}
 \affiliation{Atominstitut, Vienna University of Technology, 1020 Vienna, Austria}
\author{S. Manz}
\affiliation{Atominstitut, Vienna University of Technology, 1020
Vienna, Austria}
\author{T. Betz}
\affiliation{Atominstitut, Vienna University of Technology, 1020
Vienna, Austria}
\author{I. Bar-Joseph}
 \affiliation{Department of Condensed Matter Physics, Weizmann Institute of Science, Rehovot 76100, Israel}
 \author{T. Schumm}
\affiliation{Atominstitut, Vienna University of Technology, 1020
Vienna, Austria}
\author{J. Schmiedmayer}
 \affiliation{Atominstitut, Vienna University of Technology, 1020 Vienna, Austria}
 \affiliation{Physikalisches Institut, Universit\"at Heidelberg, 69120 Heidelberg, Germany}


\date{\today}

\begin{abstract}
We employ a combination of optical UV- and
electron-beam-lithography to create an atom chip combining
sub-micron wire structures with larger conventional wires on a
single substrate. The new multi-layer fabrication enables crossed
wire configurations, greatly enhancing the flexibility in
designing potentials for ultra cold quantum gases and
Bose-Einstein condensates. Large current densities of $>
6\times10^7$\,A/cm$^2$ and high voltages of up to 65\,V across
0.3\,$\mu$m gaps are supported by even the smallest wire
structures. We experimentally demonstrate the flexibility of the
next generation atom chip by producing Bose-Einstein condensates
in magnetic traps created by a combination of wires involving all
different fabrication methods and structure sizes.

\end{abstract}

\pacs{}

\maketitle

Manipulation of neutral atoms close to micro-structured surfaces
has become a standard technique during recent years: so called
\emph{atom chips} \cite{Fol00, For07} combine the ability to use
ultra cold atoms -- a system well suited for precise quantum
manipulation -- and the technological capabilities of micro- and
nano-fabrication. Ample techniques have been developed to trap,
cool, and detect neutral atoms in micro traps
\cite{Fol02,For07,Rei07,Liu05}. Robust quantum manipulation on the
atom chip is available, both for internal atomic states
\cite{Tre04} and the external degree of freedom \cite{Sch05b}.

\begin{figure}
 \includegraphics[width=1.0\columnwidth]{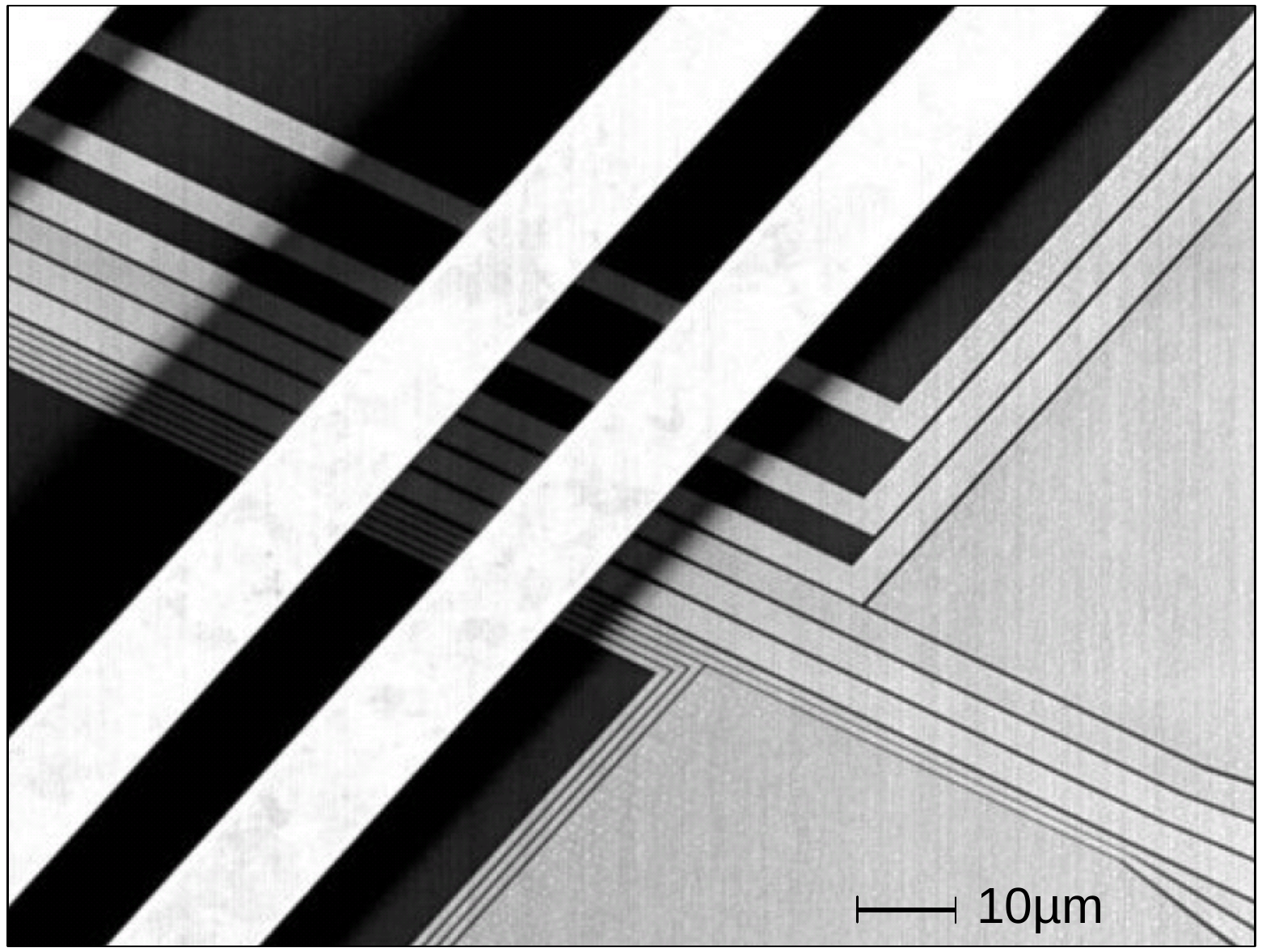}
 \caption{SEM micrograph of the central part of a multi-layer chip.
 10\,$\mu$m wide wires with a height of 1.4\,$\mu$m
 cross structures created by e-beam lithography.
 The smallest features are 300\,nm gaps between
 700\,nm wide and 140\,nm high wires.
 Electrical insulation of the two layers is provided
 by 500\,nm thick polyimide pads, visible as partially transparent layer.
 \label{fig_1}}
\end{figure}

Present atom chips are single layer devices \cite{Gro04},
sometimes combined with other structures, either macroscopic
\cite{Wil04} or built from a combination of chips fabricated on
separate substrates \cite{Gue05b,Este05}, limiting the freedom in
designing the trapping and manipulation potentials.  In this
letter we present the implementation of a multi-layer atom chip
(Fig. \ref{fig_1}) combining standard millimeter scale wires for
trapping and cooling with sub-micron structures for manipulation
on the quantum level on one single substrate. Such a chip design
offers the advantage of precise alignment of the structures given
by the inherent precision of the fabrication and the drastically
reduced spatial distance of the respective structures.

\begin{figure}
 \includegraphics[width=\columnwidth]{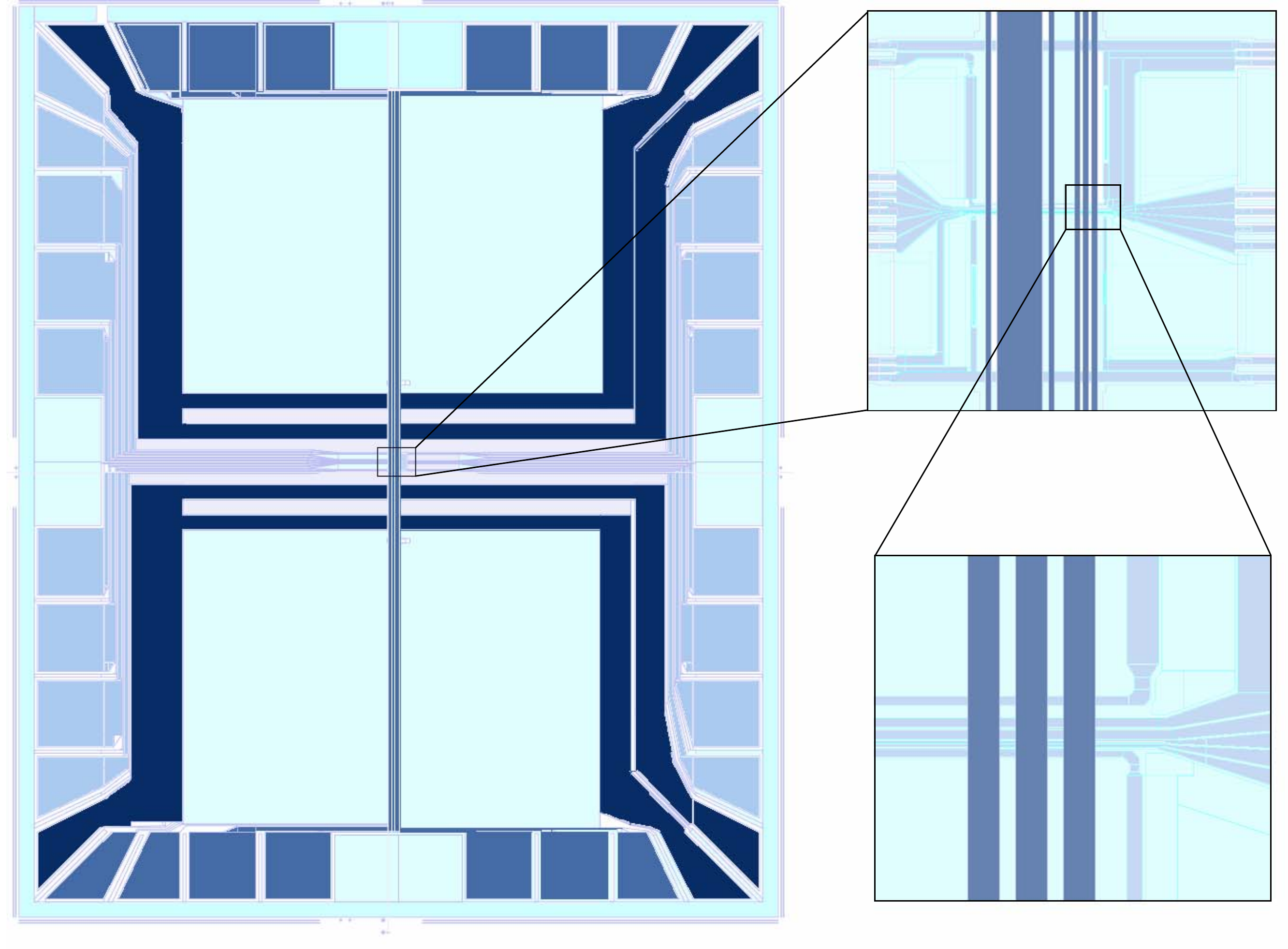}
 \caption{Layout of a multi-layer atom chip. Left:
General view of the chip, size 25$\times$30\,mm. Contact pads are
arranged around the edge of the chip. Wires for trapping of atoms
run from top to bottom (blue) on the upmost layer of the chip.
Where these wires cross structures on the ground plane, polyimide
pads provide insulation of the layers. For longitudinal
confinement of the atoms the chip contains four additional
500\,$\mu$m wide wires (dark blue) on the ground plane. Upper
right: Central part of the chip (600$\times$600\,$\mu$m$^2$),
created by e-beam lithography. Lower right: Detail of this central
section (100$\times$100\,$\mu$m$^2$), similar to the region shown
in Fig.\,\ref{fig_1}. Three 10\,$\mu$m wide wires (blue) cross
sub-micron structures (light blue, smallest features: 300\,nm wide
gaps) separated by polyimide pads.\label{fig_2}}
\end{figure}

To implement vastly different structure sizes on a single chip we
use a combination of traditional optical UV-  and electron beam
lithography. The new design consists of a standardized wire
pattern (fabricated by UV-lithography) which provides the backbone
of the atom chip with all the connections and auxiliary wires
needed for trapping, cooling and positioning the ultra cold atoms
(Fig. \ref{fig_2},\,left). This general structure is complemented
by an e-beam written part which can be custom designed for each
chip realization individually (Fig. \ref{fig_2},\,right), offering
high flexibility. A third pattern created by UV-lithography above
the e-beam layer adds chip wires to transport the atoms towards
the e-beam structures and provides additional functionality.

The main requirements for such an atom chip are the ability to
carry sizable currents of a few amps in the large structures to
create deep traps, high current densities in the small structures
for tight confinement, \cite{Fol02,For07}, to allow the
application of radio-frequency (RF) fields \cite{Sch05b,Hof06} and
the capability to tolerate sizable voltages over sub-micron gaps
to enable localized manipulation by electric fields \cite{kru03}.
To achieve this one needs a perfectly electrically insulating
layer with the capability of sufficient heat transfer from the
upper wires to the substrate, which also withstands high electric
fields. In our chip design we achieve this by separating the
different conducting layers by an electrically insulating thin
(500\,nm) polyimide layer. Conducting structures are fabricated by
thermally evaporating gold layers (thicknesses from 130\,nm to
4\,$\mu m$), which provides the best achievable wire quality
together with optimal optical reflection \cite{kru07}.

Our atom chips are fabricated on commercial 700\,$\mu$m thick
p-type Si-wafers with a 100\,nm thermal oxide layer for electrical
isolation \cite{Gro04}. We start by depositing alignment marks
which allow to superimpose the different layers with sub-micron
precision. We then fabricate the sub-micron structures in the
center 600$\times$600\,$\mu$m$^2$ of the chip
(Fig.\,\ref{fig_2},\,right). A double-layer PMMA resist (PMMA 495K
and 950K) is structured by e-beam lithography followed by
depositing first a Ti adhesion layer (10\,nm) and then the Au
layer (130\,nm), both by thermal evaporation at a pressure of
10$^{-7}$\,mbar. Lift-off in acetone supported by ultrasound then
completes fabrication of the inner sub-micron structures of the
chip.

In the next step the connections to this central part, the pads
for contacting the chip and all other larger support structures
are fabricated. We therefor employ our standard process for high
quality atom chip structures \cite{Gro04}: the image reversal
resist AZ5214E is structured by traditional UV contact lithography
followed by thermal evaporation of Ti (20\,nm) and Au (400\,nm in
this specific example). The structures are created again by
lift-off in acetone.

\begin{figure}
\includegraphics[width=\columnwidth]{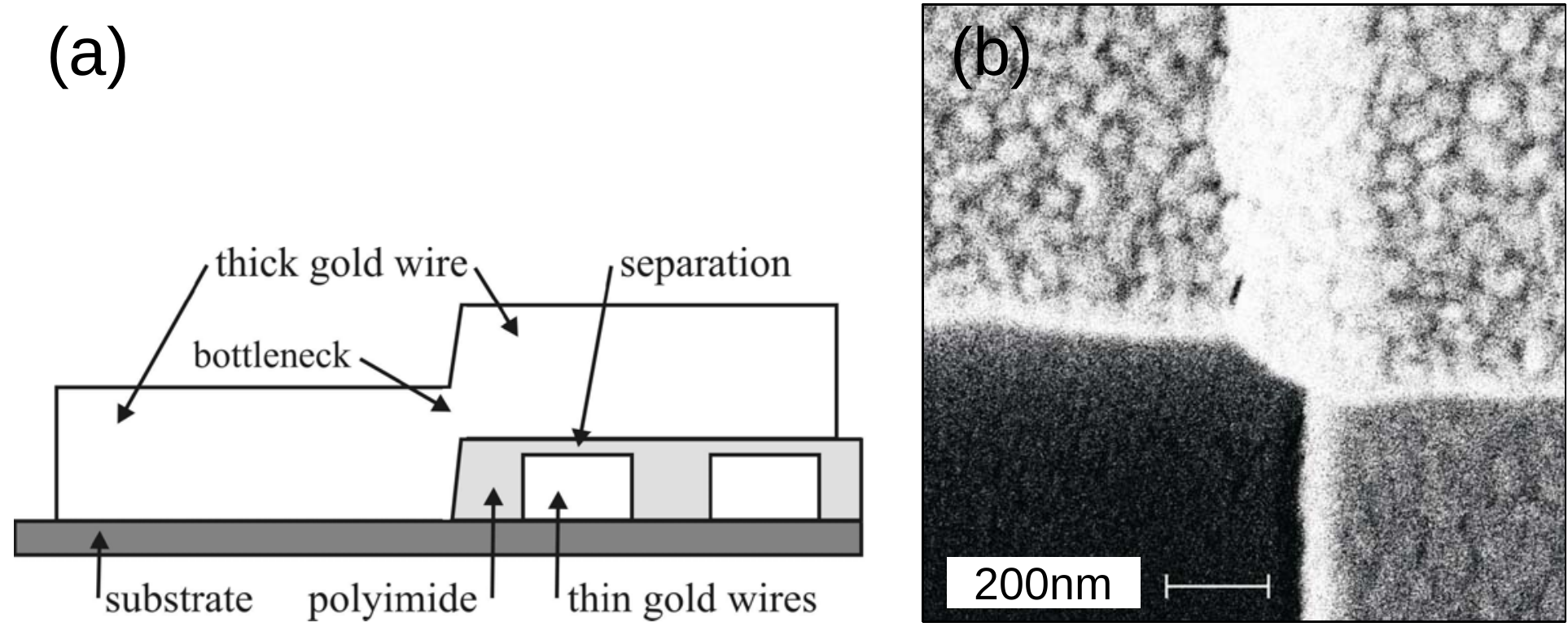}
\caption{(a): Cross section scheme of a multi-layer area (not to
scale). The step in the upper gold wire causes a bottleneck of
reduced cross section. (b): SEM top view of the step. The wire
runs from left to right in the upper half of the picture. In the
lower right part the polyimide pad running from top to bottom is
visible.\label{fig_3}}
\end{figure}

We then prepare the insulation which will support the crossing
structures: Polyimide (Durimide (R) 7505) is spun onto the chip
and structured by UV-lithography to cover only the regions where
conducting structures will cross. The insulation layer is then
thinned in an ozonator to about 500\,nm and cured. This layer
thickness proved to be sufficient to insulate the two conducting
planes while providing good heat transfer and keeping the step
height the top layer wires have to surmount to a minimum
(Fig.\,\ref{fig_3}).

The wires in the upper plane crossing the polyimide insulation
pads are fabricated again by our standard UV-lithography process
\cite{Gro04}. To enable high currents in these wires they can be
evaporated to a height of up to 4\,$\mu$m. This also reduces the
bottleneck created by the step onto the polyimide pads
(Fig.\,\ref{fig_3}).

The current characteristics of the various atom chip wires are
tested by a four-point measurement, monitoring heating via
resistance increase with time. The cold resistance of the
structures ranges from 100\,$\Omega$ to 300\,$\Omega$ for
e-beam-written wires, and 4.5\,$\Omega$ to 40\,$\Omega$ for the
larger structures. Similar to the situation in the actual atom
chip experiments the measurements were carried out in a pulsed
manner with a 10\,s relaxation time.

\begin{figure}[t]
\includegraphics[width=\columnwidth]{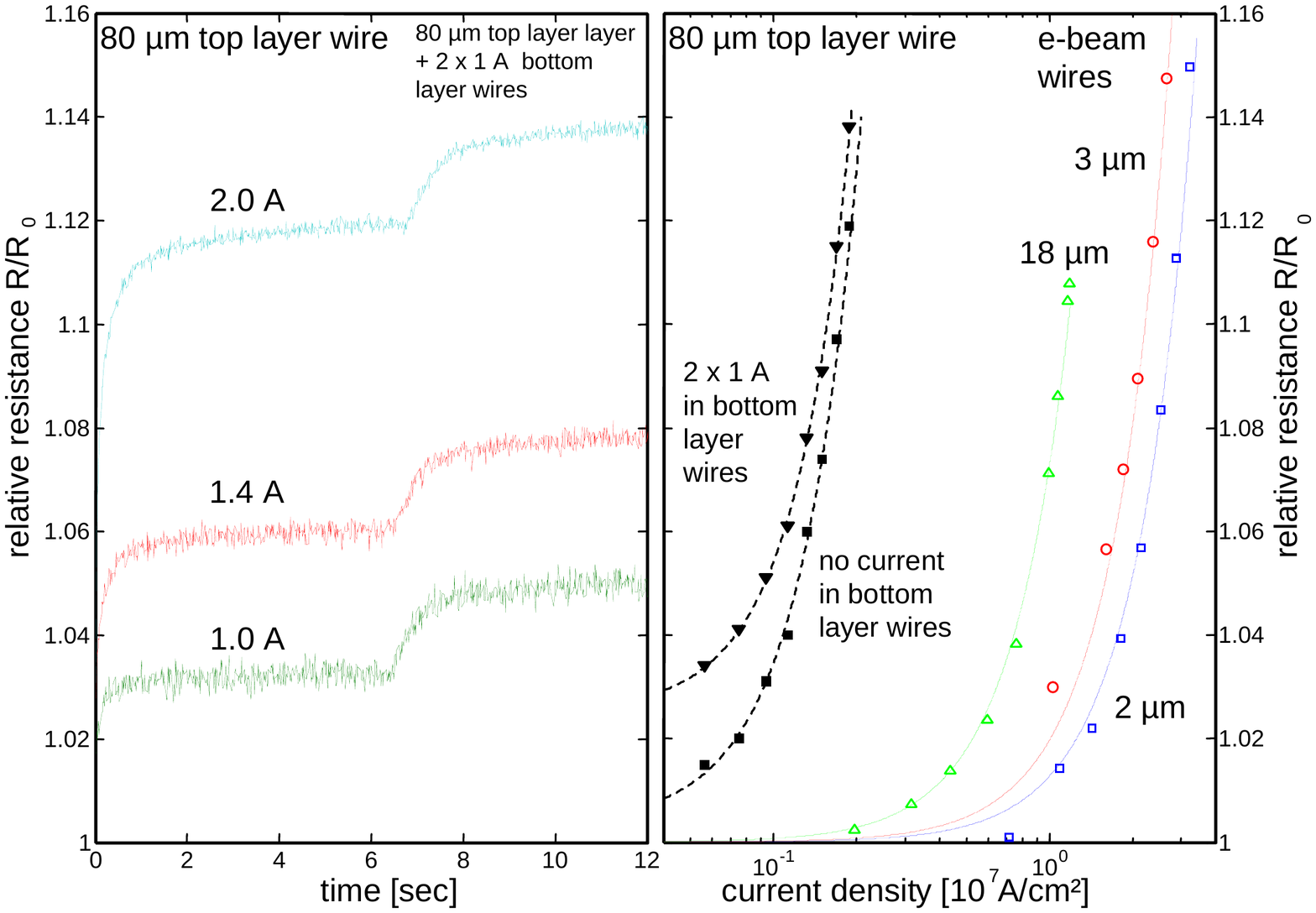}
\caption{ \emph{left:} Temperature evolution of a 80\,$\mu$m top
layer trapping wire for different applied currents. After
$\approx$\,6\,s, an additional current of 1\,A is send through two
500\,$\mu$m bottom layer confinement wires, as in the experiment
shown in Fig. \ref{fig_6}. \emph{right:} Temperature evolution for
different current densities in various chip wires. Solid lines are
theoretical predictions according to a simple dissipation model
which applies to bottom layer e-beam wires in direct contact with
the substrate \cite{Gro04}. Reduced heat dissipation reduces the
current density for top layer wires, currents in the bottom layer
wires lead to additional heating (dashed lines to guide the eye).
\label{fig_4} \label{fig_5}}
\end{figure}

For surface mounted wires the results are similar to what was
found in our previous study on single layer atom chips
\cite{Gro04}: the heating process of  structures shows two
different time scales. The flat wires ($\sim$\,100\,nm) first heat
up on a fast time scale of 100 ns after switching on the current,
leading to a corresponding increase in wire resistance. The
corresponding time scale for higher structures ($> \,1 \mu$m) is
considerably longer, in the range of 1\,$\mu$s. On a longer
timescale a slow heating process is observed over the full
duration of the current pulse (Fig.\,\ref{fig_4}). In accordance
with the model from \cite{Gro04} the highest current densities
were tolerated by the wires of smallest cross section, see
Fig.\,\ref{fig_5}. A 700\,nm wide and 140\,nm high wire carried
currents of up to 60\,mA over a maximum of 10\,s, corresponding to
a current density of $6 \times10^7$\,A/cm$^2$.

For the wires crossing the polyimide insolation pads we observe a
similar behavior with a reduced maximal current due to the
decreased thermal conductivity to the substrate. Nevertheless
large current densities of above $3 \times 10^6$\,A/cm$^2$ ($2
\times 10^6$\,A/cm$^2$) can be supported by a 10\,$\mu$m
(80\,$\mu$m) wide wire (Fig.\,\ref{fig_4}).  DC Currents of up to
2.3\,A are sustained by the 80\,$\mu$m wide and 1.4\,$\mu$m high
wire.

A new crucial test for the double-layer chip structures is to
which extend a current or voltage in the wires in the bottom layer
influences the current limits in the wires above. We study this by
measuring the temperature evolution of a 80\,$\mu$m top layer wire
after switching on additional 500\,$\mu$m wide 400\,nm high
confinement wires in the ground plane. A clear increase in
resistance and hence heating of the top wires depending on the
current density in the lower wires can be observed (Fig.
\ref{fig_5}). For the largest currents in the bottom wires the
maximum current sustained in the top wires can be reduced by a
factor of two.

The combination of sub-micron wires connected to larger structures
causes non-uniform heat dissipation and can lead to delicate
failure modes. The resistivity change is not a firm indication any
more, as it was in the single layer chips \cite{Gro04}. The
different parts of the wire contribute differently (locally) to
heating, and a simple rule of how much resistivity increase can
safely be tolerated can not be determined. Hence, current limits
have to be found for each individual wire category. For example
the sub-micron wires in the central part tend to burn for an
increase of $R$/$R_o$ by more than 20\,\%, whereas the larger
wires in a single layer chip easily support an increase by
100\,\%.

In voltage tests, even the smallest wire structures allowed to
apply sizable voltages of $>$ 65\,V without breakdown, the
insulation layer was not effected. This allows to apply electric
field of over 200\,kV/mm over the 300\,nm gaps between e-beam
wires, creating extremely steep and localized potentials
\cite{kru03}.

Our multi-layer atom chip combines many functionalities in a
single device: the large top layer wires (10 to 80\,$\mu$m width,
1.4\,$\mu$m height) can carry sizable currents of above 2\,A to
create the magnetic traps for trapping, cooling and positioning
ultra cold atom clouds or Bose-Einstein condensates (BECs)
together with strong RF oscillating fields for dressed-state
potentials \cite{Sch05b,Hof06}. Large bottom layer confinement
wires equally enable high currents and allow adjustment of trap
aspect ratio over many orders of magnitudes. The small e-beam
wires allow micron size structuring of the potentials for
manipulation of the trapped atoms on a scale where tunnelling and
coupling between traps can be studied.

\begin{figure}[t]
\includegraphics[width=\columnwidth]{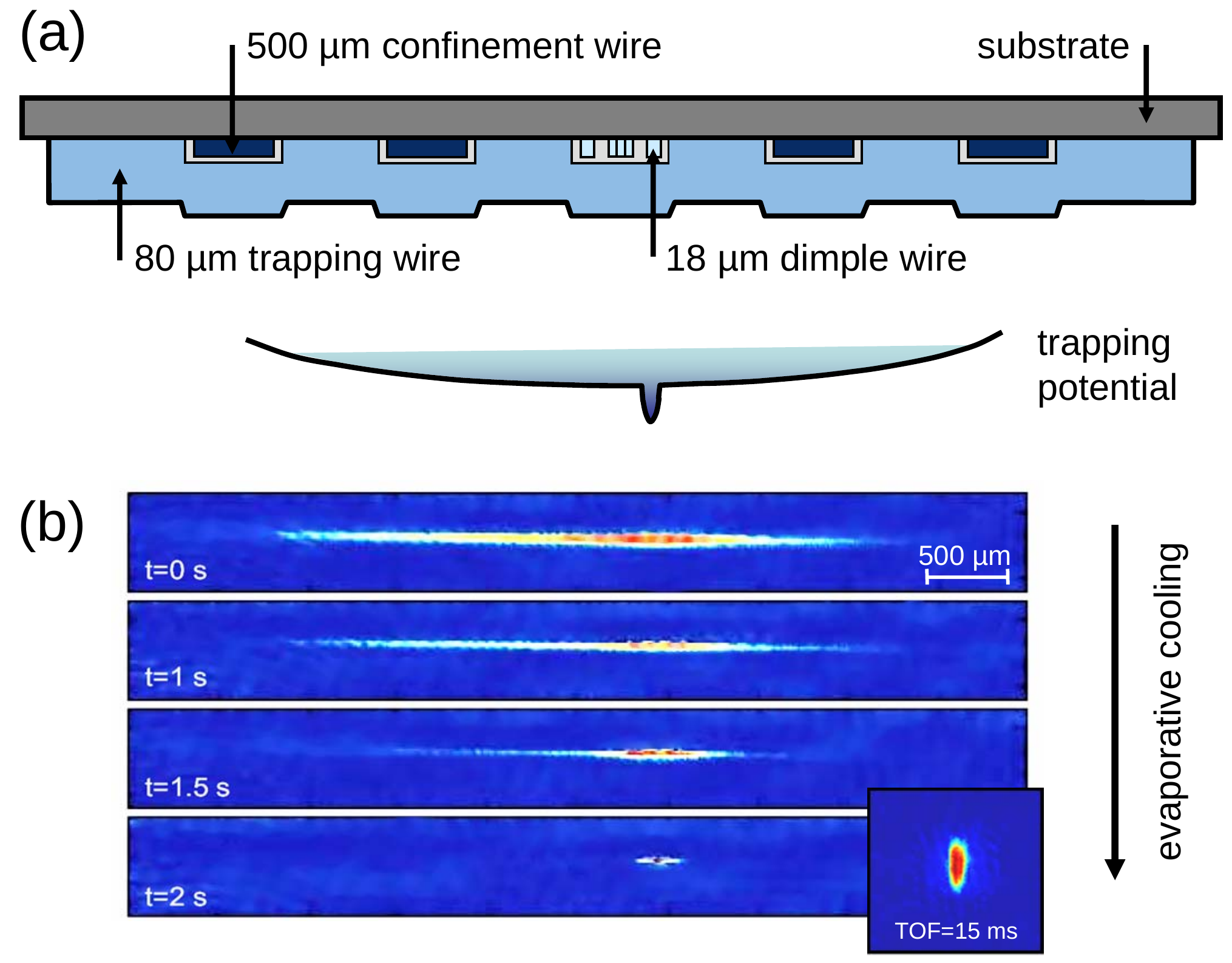}
\caption{(a): Magnetic trapping potential created by the combined
fields of the outer 500\,$\mu$m bottom layer confinement wires
(dark blue), the main 80\,$\mu$m top layer trapping wire (blue)
and a 10\,$\mu$m e-beam fabricated dimple wire (light blue),
creating a local potential minimum. (b) In-situ absorption images
of an atom cloud, evaporatively cooled in the combined potential.
Inset: time-of-flight (TOF) absorption image after 15\,ms free
expansion, clearly indicating Bose-Einstein
condensation.\label{fig_6}}
\end{figure}

We experimentally demonstrate the double-layer atom chip
flexibility by creating Bose-Einstein condensates of Rubidium
atoms in a magnetic trapping potential created by a combination of
all structures described above (Fig.\,\ref{fig_6}a). Starting
point is a reflection magneto-optical trap, using the high quality
gold surfaces of the atom chip as a mirror. Macroscopic copper
wire structures below the chip create the magnetic fields
necessary for laser cooling, initial magnetic trapping and
transport to the atom chip \cite{Wil04}. Atoms are then loaded
into a chip trap combining magnetic fields of a top layer
80\,$\mu$m trapping wire and two 500\,$\mu$m bottom layer
confinement wires in series, each carrying 1\,A. Sending 40\,mA
through a 18\,$\mu$m bottom layer e-beam wire locally lowers the
potential, creating an adjustable magnetic dimple \cite{Sta98}.
Efficient and robust Bose condensation is achieved by forced RF
evaporation in the combined trap as shown in Fig.\,\ref{fig_6}b.

To summarize, we have presented fabrication, characterization and
implementation of an atom chip combining structures created with
traditional UV- and e-beam lithography in a multi-layer geometry.
With this concept we integrated wires tolerating extreme current
densities and electric fields with established structures for
robust trapping of cold atoms in a single device. We have shown
that the temperature evolution of the surface mounted structures
agrees with a simple dissipation model. In addition, the influence
of currents in the ground plane wires on the resistance of wires
crossing these structures was analyzed. We experimentally
demonstrate the atom chip flexibility by producing BECs in a
potential involving all of the major wire structures.

We thank O. Raslin, Braun Center for Submicron Research at the
Weizmann Institute of Science, for help in the fabrication. This
work was supported by the European Union, contract Nos.
IST-2001-38863 (ACQP), MRTN-CT-2003-505032 (Atomchips), Integrated
Project FET/QIPC 'SCALA', the Deutsche Forschungsgemeinschaft, the
Austrian Science Fund FWF Project P20372, and the Wittgenstein
Prize of the Austrian Science Fund FWF.


\end{document}